# Field-Induced Tunneling Ionization and Terahertz-Driven Electron Dynamics in Liquid Water

Ahmed Ghalgaoui, Lisa-Marie Koll, Bernd Schütte, Benjamin P. Fingerhut, Klaus Reimann, Michael Woerner, Thomas Elsaesser*

*Max-Born-Institut für Nichtlineare Optik und Kurzzeitspektroskopie,*

*Berlin 12489, Germany*

*Corresponding author: elsasser@mbi-berlin.de

Abstract

Liquid water at ambient temperature displays ultrafast molecular motions and concomitant fluctuations of very strong electric fields originating from the dipolar $H_2O$ molecules. We show that such random intermolecular fields induce tunnel ionization of water molecules, which becomes irreversible if an external terahertz (THz) pulse imposes an additional directed electric field on the liquid. Time-resolved nonlinear THz spectroscopy maps charge separation, transport and localization of the released electrons on a few-picosecond time scale. The highly polarizable localized electrons modify the THz absorption spectrum and refractive index of water, a manifestation of a highly nonlinear response. Our results demonstrate how the interplay of local electric field fluctuations and external electric fields allows for steering charge dynamics and dielectric properties in aqueous systems.

TOC Graphic

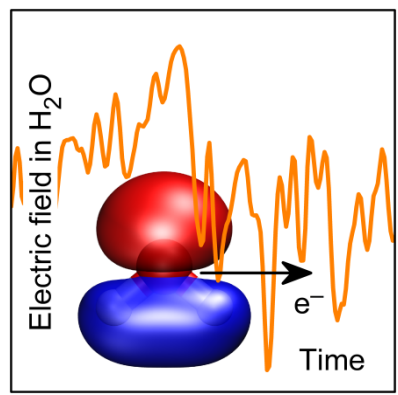

Keywords: water, solvated electron, tunneling ionization, nonlinear terahertz spectroscopy, low-frequency dielectric function.

Electrons and ions solvated in water have a strong impact on the physical and chemical properties of the liquid. The aqueous or hydrated electron, a free electron hosted by an environment of polar $H_2O$ molecules, represents a prototypical system. Both its stationary properties and nonequilibrium dynamics are the subject of current research.[1-5] After injection into the liquid by electron irradiation or by ionization of a water molecule, a free electron undergoes a sequence of femto- to picosecond processes. It eventually localizes in a self-consistent potential well defined by the water environment.[6-11] While electron relaxation and optical properties of the confined electron have been analyzed in some detail, the impact of strong external electric fields on water ionization, electron transport, and electron solvation has remained unaddressed. Moreover, the influence of electron solvation on the excitation spectrum of the liquid in the THz range, in particular the important intermolecular and collective degrees of freedom, is not known.

A strong electric field can induce tunnel ionization of a water molecule, by which an electron leaves its binding orbital (e.g., the HOMO or HOMO-1, Figure 1a,b) and moves into a spatially delocalized state in the liquid environment.[11-14] This mechanism requires a field-induced distortion of the molecule's binding potential for generating a tunneling barrier narrow enough to allow the bound electron to escape (Figure 1c). A water molecule embedded in the liquid is subject to the fluctuating electric field from the neighboring water molecules. This field reaches amplitudes on the order of 100 MV/cm and displays major frequency components in the range from 20 GHz to 20 THz (Figure 1a,b, ref 15,16). While the field strength from the environment is sufficient for inducing a narrow tunneling barrier, the macroscopic rate of successful ionization events is extremely small. This behavior is due to a lack of persistent charge separation in the temporally and directionally fluctuating external field, i.e., the electron cannot undergo a directed motion away from the parent ion. As a result, charge recombination at the ionization site prevails. Details of this scenario are discussed in the Supplementary Information (SI).

This behavior should change dramatically whenever an external directed electric field of sufficient strength is superimposed on the fluctuating field from the water molecules. In this way, a separation of the electron from the parent ion and electron transport along the external field direction become feasible (Figure 1d). For an irreversible charge separation, the electron needs to acquire an amount of energy in the external electric field that is on the order of the ionization potential of the water molecule. Here, we demonstrate the occurrence of field ionization, separation of the electron from the parent water ion and subsequent localization in presence of a strong external THz field with frequencies in the range of the fluctuating intermolecular fields. Electron dynamics are followed via the time-resolved THz dielectric response of the liquid by two-dimensional terahertz (2D-THz) spectroscopy.[17]

In the experiments, THz pulses of an electric-field amplitude between 100 kV/cm and 2 MV/cm interact with a free-standing liquid-water jet (thickness 50 µm) at ambient temperature. The spectra of the THz pulses span more than an octave in frequency with a maximum around 0.7 THz. THz generation and experimental methods are described in the SI. To verify the generation of solvated electrons upon THz excitation, we performed pump-probe experiments in which a THz pulse with a peak value $E_{A,P}$ = 1.9 MV/cm of the transmitted electric field excites the water jet. The resulting absorption change of the sample in the near-infrared is then probed with a femtosecond white-light continuum. In Figure 1e, the transient absorption recorded at a delay time $\tau$ = 300 ps is plotted as a function of probe wavelength $\lambda_{pr}$ (symbols). Within the experimental accuracy, the data points are in agreement with the scaled absorption spectrum of solvated electrons (blue line, ref 18), i.e., the strong-field THz excitation definitely creates solvated electrons. Using the absolute molar extinction coefficient $\varepsilon$ = 19000 M$^{-1}$cm$^{-1}$ at 800 nm, one derives a concentration of THz-generated solvated electrons of $c_e \approx 2 \times 10^{-5}$ M = (20±3) µM for $E_{A,P}$ = 1.9 MV/cm, and $c_e$ = (5±1) µM for $E_{A,P}$ = 500 kV/cm as observed in a second measurement.

In the 2D-THz experiments, two phase-locked THz pulses A and B with peak electric fields of up to 500 kV/cm interact in a collinear geometry with the water jet (Figure 1f). In Figure 2d, the electric-field transients of the stronger pulse $E_A(t, \tau = 0)$ (red line) and the weaker pulse $E_B(t)$ (orange line) transmitted through the sample are plotted as a function of real time $t$. In a 2D-THz measurement, the nonlinear THz response is mapped via the electric field $E_{NL}(t, \tau) = E_{AB}(t, \tau) - E_A(t, \tau) - E_B(t)$ emitted by the water sample.[17] Here, $E_{AB}(t, \tau)$ is the transmitted electric field after interaction with both pulses A and B separated by the delay time $\tau$, and $E_A(t, \tau)$, $E_B(t)$ are the electric fields transmitted after interaction with pulse A or pulse B only. The delay time $\tau$ is measured relative to the maximum of pulse B, i.e., for $\tau > 0$ pulse A interacts with the sample before pulse B.

In Figure 2a we show a contour plot of the electric field $E_{AB}(t, \tau)$ as a function of real time $t$ and delay time $\tau$. Figure 2b displays the nonlinearly emitted electric field $E_{NL}(t, \tau)$. For zero delay between pulses A and B, $E_{NL}(t, \tau = 0)$ reveals a high signal amplitude of up to 100 kV/cm, corresponding to one third of the maximum field of pulse B. A two-dimensional Fourier transform of $E_{NL}(t, \tau)$ along $t$ and $\tau$ generates the frequency-domain signal $E_{NL}(\nu_t, \nu_\tau)$, plotted in Figure 2c as a function of detection frequency $\nu_t$ and excitation frequency $\nu_\tau$. The contours represent an A-pump–B-probe signal, which dominates the nonlinear response. Other nonlinear signals, in particular third-order photon echoes, are absent.

Measurements with a different peak field strength $E_{A,P}$ of the transmitted pulse A reveal a sharp threshold $E_{A,P} \geq 250$ kV/cm for the occurrence of the nonlinear THz response. In Figure 2d,e, we present transients measured above and below this threshold. The nonlinear signal field $E_{NL}(t, \tau = 7 \text{ ps})$ at a delay time $\tau = 7$ ps is plotted as a function of real time $t$ (blue lines), together with the probe field $E_B(t)$ (orange lines) and $E_{B,T}(t, \tau = 7 \text{ ps}) = E_B(t) + E_{NL}(t, \tau = 7 \text{ ps})$ (black dashed lines). For $E_{A,P} \leq 200$ kV/cm (Figure 2e), there is no nonlinear response of the sample within our signal-to-noise ratio, i.e., $E_{NL}(t, \tau = 7 \text{ ps}) = 0$ and $E_{B,T}(t, \tau = 7 \text{ ps}) = E_B(t)$. For $E_{A,P} = 500$ kV/cm (Figure 2d), a strong nonlinear response is observed with a maximum nonlinear signal field on the order of 100 kV/cm. The transient $E_{B,T}(t, \tau)$ exhibits a larger amplitude than $E_B(t)$ and a slight phase shift towards smaller real times, both arising from a nonlinear change of the complex refractive index of the water sample.

The evolution of the nonlinear signal field as a function of the delay $\tau$ between pulses A and B is shown in Figure 2f where $E_{NL}(t = -0.2$ ps, $\tau)$ (red lines) and $E_{NL}(t = 0.2$ ps, $\tau)$ (blue lines) from two different measurements are plotted as a function of $\tau$. The curves exhibit an initial fast kinetics within the temporal overlap of pulses A and B of approximately 2 ps, followed by a slower gradual signal change on a time scale up to some 10 ps. At even longer delays $\tau$ up to 40 ps, the signal fields $E_{NL}(t = \pm 0.2$ ps, $\tau)$ remain essentially constant.

For an in-depth analysis of the 2D-THz data, we perform a Fourier transform of the time-dependent THz fields along $t$ into the frequency domain. The spectrum $|E_{B,T}(\nu)|$ of the probe pulse B transmitted after THz excitation (symbols in Figure 3a) displays a higher amplitude than the spectrum transmitted without excitation (black line), giving evidence of an absorption decrease upon excitation. To gain quantitative insight, the nonlinear response is described as a pump-induced change of the complex refractive index $n(\nu) = n_{re}(\nu) + i\, n_{im}(\nu)$ as probed by pulse B. In Figure 3b, the real and imaginary parts of $n(\nu)$ of water in equilibrium, i.e., without THz excitation, are plotted in the frequency range from 0 to 2 THz (black lines (1,1'), ref 19,20). After THz excitation by pulse A, the complex ratio of the transmission of the probe field with $[E_{B,T}(\nu)]$ and without $[E_B(\nu)]$ THz pump is given by $E_{B,T}(\nu)/E_B(\nu) = \exp\{i\, 2\pi\nu[n'(\nu)-n(\nu)]d/c_0\}$ with the modified refractive index $n'(\nu)$, the sample thickness $d$, and the vacuum velocity of light $c_0$. This expression allows for deriving $n_{re}'(\nu)$ and $n_{im}'(\nu)$ from the measured electric fields $E_{B,T}(\nu)$ and $E_B(\nu)$. In Figure 3b, $n_{re}'(\nu)$ (open symbols) and $n_{im}'(\nu)$ (full symbols) are presented. Both quantities exhibit a marked decrease by ~10% compared to the stationary equilibrium values $n_{re}(\nu)$ and $n_{im}(\nu)$ (black lines (1,1')). The smaller value of $n_{im}'(\nu)$ corresponds to a reduced THz absorption of the sample (cf. Figure 3a).

The weak absorption of water in the frequency range of the THz pulses results in the excitation of less than $2\times10^{-3}$ of the water molecules in the interacting sample volume (cf. SI). Thus, any contribution of a nonlinear change of resonant THz absorption to the observed THz response is negligible. Because of the very small amount of energy the THz pulses deposit in the sample, heating effects can be neglected as well. Potential changes of the refractive index due to the THz Kerr effect[21,22] were investigated in separate experiments presented in the SI. The Kerr-induced change of refractive index is several orders of magnitude smaller than the 10% changes of refractive index shown in Figure 3b and, thus, can be safely neglected.

To benchmark the nonlinear response observed in the 2D-THz experiments, we performed reference experiments in which water molecules were ionized by multi-photon excitation with femtosecond near-infrared pulses centered at 800 nm. This mechanism generates free electrons, which are solvated in the water environment on a time scale of a few picoseconds.[6-11] The resulting change of THz transmission was recorded with a probe pulse $E_B(t, \tau)$ as a function of $t$ and $\tau$ (cf. SI). In Figure 3f,g, transient THz absorption spectra (symbols) are presented for energies of the 800 nm pump pulse of $W_{pu} = 0.94$ mJ and $W_{pu} = 1.32$ mJ, corresponding to respective electron concentrations $c_e = 24$ μM and $c_e = 85$ μM. At the lower $c_e$ (Figure 3f), one observes an absorption decrease, while an increase of absorption arises for the higher $c_e$ (Figure 3g, further results in the SI).

The strong changes in both the real and imaginary part of the refractive index upon THz excitation are a hallmark of solvated electrons, generated by field ionization of water molecules and charge separation in the strong external THz field. Field ionization requires an electric field in the 100 MV/cm range to polarize the water molecule and generate a finite narrow potential barrier (Figure 1c), through which an electron tunnels from the highest occupied molecular orbitals $1b_1$ (HOMO) or $3a_1$ (HOMO-1) into continuum states.[12,13] The stochastic random electric field from the water environment reaches peak amplitudes on the order of 100 MV/cm (cf. Figure 1a,b), high enough for inducing electron tunneling. A successful, i.e., persistent ionization event requires on top the spatial separation of the released electron from the parent $H_2O^+$ ion (Figure 1d), i.e., charge transport in real space, to suppress on-site charge recombination. In our experiments, the electron is strongly accelerated during a half cycle of the directed electric field of the THz pump pulse. As a result, it moves away from the parent ion, and eventually localizes at a new site in the liquid. To make an irreversible separation of parent ion and electron happen, the THz field has to be present at the instant of the ionization, which happens on a time scale much shorter than the THz oscillation period. The THz field has a negligible impact on ionization which occurs without external field as well (autoionization). In the latter case, however, the spatial separation of parent ion and electron is insufficient and charge recombination on a (sub)-100 fs time scale prevails, as addressed in detail in the SI.

A quantum-mechanical description of this scenario gives insight in the temporal evolution of the electron wavefunction and the interaction with both the random electric field of the liquid and the external THz field. As discussed in the SI with the help of a one-dimensional model, the released electron represents a quantum wave packet of nanometer size, i.e., much larger than a water molecule and extending over several water layers around the ionized molecule. The nanometer extension of the wave packet results in a strong averaging and a reduction of the electric force the random water environment exerts on the released electron. Nevertheless, the fluctuating electric field with an initial correlation decay on a 20 fs time scale (cf. Fig. S7a and ref 8) induces decoherence of the quantum wave packet and, thus, a decoupling from the wave function of the parent ion. Under the influence of the fluctuating force, an electron wave packet of a spatial extension $\Delta x$ experiences a decoherence rate proportional[23] to $(\Delta x)^2$. This mechanism leads to a sub-10 fs decoherence time (cf. SI), in line with quantum-classical simulations of solvated electrons in water.[24]

The spatial separation from its parent ion requires the electron to take up a kinetic or ponderomotive energy[14] in the THz field which is larger than the ionization energy $W_I \approx 11$ eV (ref 25) of the water molecule. The maximum ponderomotive energy is given by $U = e^2 E^2/(8\pi^2 m_e \nu^2)$, when an electron (charge -$e$, mass $m_e$) is accelerated by an electric field with amplitude $E$ and frequency $\nu$. With a frequency $\nu = 0.7$ THz, the peak frequency of our THz pulses, one derives a threshold electric field of $E = 500$ kV/cm from the equation $U = W_I$. This field represents the local field required in the liquid near a water molecule. The local field is higher than the applied external field by the Clausius-Mossotti factor $(\varepsilon + 2)/3 = 2$, using $\varepsilon = 4$ as the relative dielectric constant of water at the frequency $\nu = 0.7$ THz. Thus, to achieve a ponderomotive energy of $U = W_I = 11$ eV, one needs an external threshold field with an amplitude of 250 kV/cm, which agrees very well with our experimental results.

The irreversible eventual localization of the electron at a new site in the water environment is connected with a spatial shrinking of the wave packet and picosecond energy relaxation to form a solvated (or hydrated) electron some 100 nm away from the ionization site.[6,7] The solvated electron is a highly polarizable entity with a strong impact on the dielectric function or refractive index of the liquid. In our experiments, the latter is monitored by the THz probe pulse.

A THz pump pulse with $E_{A,P} = 500$ kV/cm generates a concentration of solvated electrons of $c_e = 5$ μM or a volume density $N_e = 3\times10^{15}$ cm$^{-3}$. The average distance between solvated electrons of this concentration is on the order of 70 nm, rendering interactions between them negligible. Each solvated electron displays a frequency-dependent electric polarizability $\alpha_{el}(\nu) = -e^2 / (\varepsilon_0 m_e[(2\pi\nu)^2 + i2\pi\nu\gamma])$ ($\varepsilon_0$: vacuum permittivity, $\gamma \approx 1$ ps$^{-1}$ damping rate) and, thus, contributes to the local electric field in the liquid. The localized character of the solvated electrons makes the Clausius-Mossotti concept applicable for incorporating local-field effects in dielectric response theory.[26,27] The total dielectric function $\varepsilon(\nu)$ of liquid water plus solvated electrons is given by the following expression:

$$3\frac{\varepsilon(\nu) - 1}{\varepsilon(\nu) + 2} = 3\frac{\varepsilon_{water}(\nu) - 1}{\varepsilon_{water}(\nu) + 2} + N_e \alpha_{el}(\nu) \qquad (1)$$

Here, the much heavier positive ions are not included because of their negligible polarizability. The black lines (1,1') in Figure 3b show the real part (solid line) and the imaginary part (dashed lines) of the refractive index $n(\nu) = [\varepsilon_{water}(\nu)]^{0.5}$ of neat water in the frequency range from 0 to 2 THz. The red lines (2,2') represent the results for an electron concentration of $c_e$ = 5 µM calculated from eq. 1, in good agreement with our experimental data (symbols). Theory and experiment demonstrate a reduction of both the real and the imaginary part of the complex refractive index by approximately 10%. The measured amplitude change in the spectrum of the transmitted probe pulse B (Figure 3a) agrees very well with the theory result (red line). A further hallmark of solvated electrons is the build-up of the nonlinear signal field $E_{NL}(t = \pm 0.2$ ps, $\tau)$ (Figure 2f), which extends over 5 to 6 ps due to the relaxation of the electrons in their local potential minima. This time scale agrees with extensive studies of electron solvation in the literature.[6-11]

To validate the impact of solvated electrons on the dielectric spectra of water in a broader context, the experiments with 800 nm pump and THz probe pulses were analyzed with the same model. As shown in Figure 3f,g, generation of solvated electrons after 800 nm multi-photon ionization of water leads to pronounced negative THz absorption changes at low electron concentrations $c_{el}$ (pump energy $W_{pu}$ = 0.94 mJ) and to positive absorption changes at $W_{pu}$ = 1.32 mJ. Based on eq 1, we calculated the frequency- and concentration-dependent THz absorption coefficient $\alpha_{abs}(\nu,c_e) = 4\pi(\nu/c_0)n_{im}(\nu)$ (Figure 3d) and differential absorption coefficient $\Delta\alpha_{abs} = \alpha_{abs}(\nu,c_e) - \alpha_{abs}(\nu,0)$ (Figure 3e) with $\alpha_{abs}(\nu,0)$ shown in Figure 3c. The dashed lines in Figure 3f and Figure 3g show the differential absorption spectra for $c_e$ = 24 µM and 85 µM, respectively. Both the spectral position of the maximum and the sign change of differential absorption with increasing concentration are well reproduced by the calculations.

In conclusion, our results establish high-field tunnel ionization of water, charge separation and subsequent electron localization as predominant mechanisms behind the nonlinear low-frequency dielectric response of water. Such processes are driven by the interplay of strong fluctuating intermolecular electric fields and a much weaker directed external THz electric field. Beyond the insight in the nonlinear THz response of water, the present work establishes a concept for enhanced charge separation and manipulation in aqueous systems with THz fields.

*Acknowledgments.* This research has received funding from the European Research Council (ERC) under the European Union's Horizon 2020 research and innovation program (grant agreements No. 833365 and No. 802817). B. P. F. acknowledges support by the DFG within the Emmy-Noether Program (Grant No. FI 2034/1-1). M. W. acknowledges support by the DFG (Grant No. WO 558/14-1).

*Supporting Information Available.* Materials and methods, experimental results, theoretical methods and results, Figures S1 to S8.

**Figures**

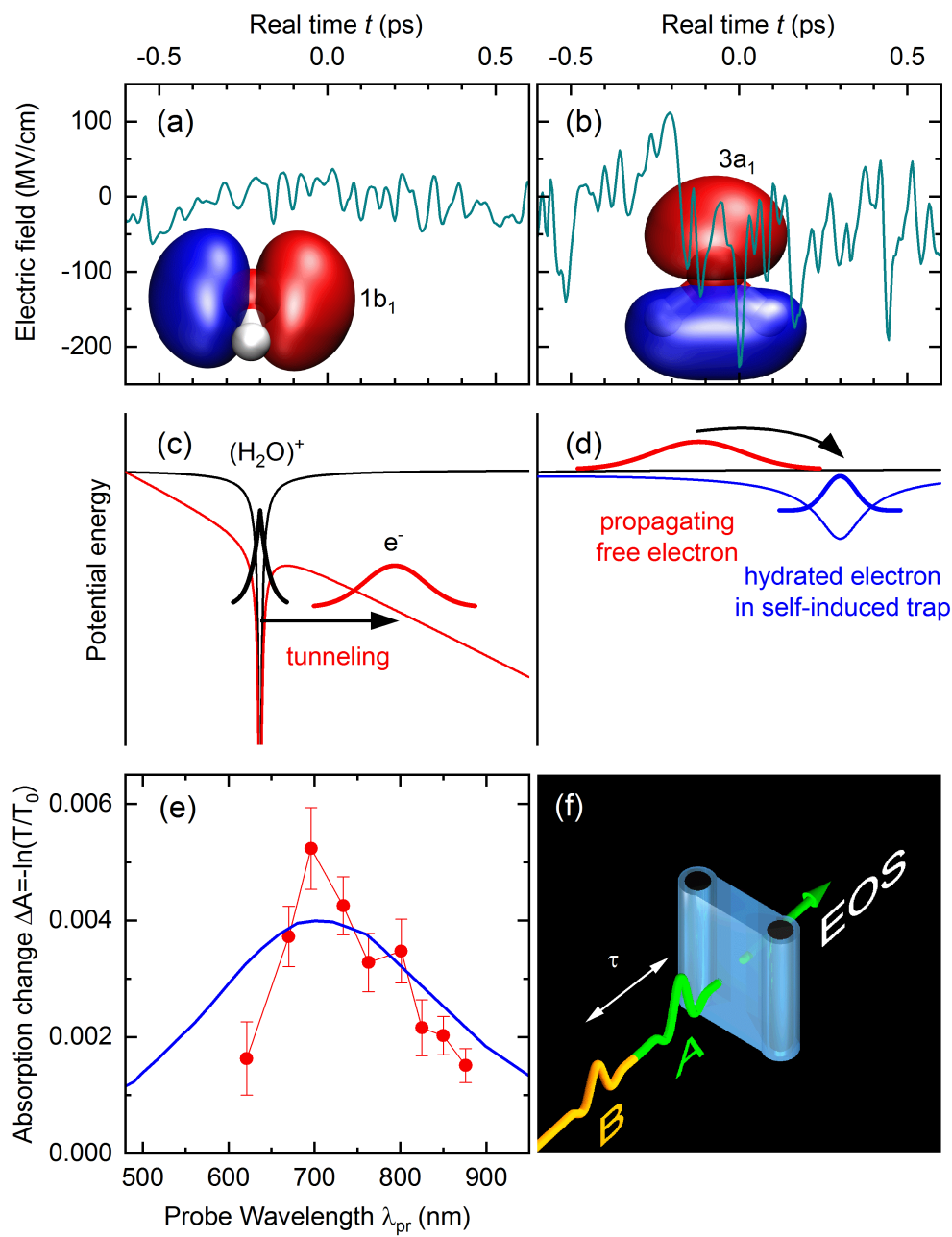

**Figure 1**. Field-induced electron dynamics in water and nonlinear two-dimensional terahertz (2D-THz) spectroscopy. (a, b) Fluctuating electric fields in liquid water as calculated from a molecular dynamics simulation for electrons at the position of the two highest occupied orbitals $1b_1$ (HOMO) and $3a_1$ (HOMO-1). (c) Schematic diagram of field-induced electron tunneling. (d) Schematic of electron propagation and localization. (e) Near-infrared absorption change $\Delta A = -\ln(T/T_0)$ of the water sample at a delay time $\tau = 300$ ps after interaction with a THz pump pulse of a peak field $E_{A,P} = 1.9$ MV/cm transmitted through the sample (symbols, $T$, $T_0$: sample transmission with and without excitation). Blue line: Scaled absorption spectrum of solvated electrons taken from ref 18. (f) Schematic of the 2D-THz experiment with the two phase-locked THz pulses A and B of delay $\tau$, the water jet (blue) in a transmission geometry, and the path towards the electrooptic sampling (EOS) detector.

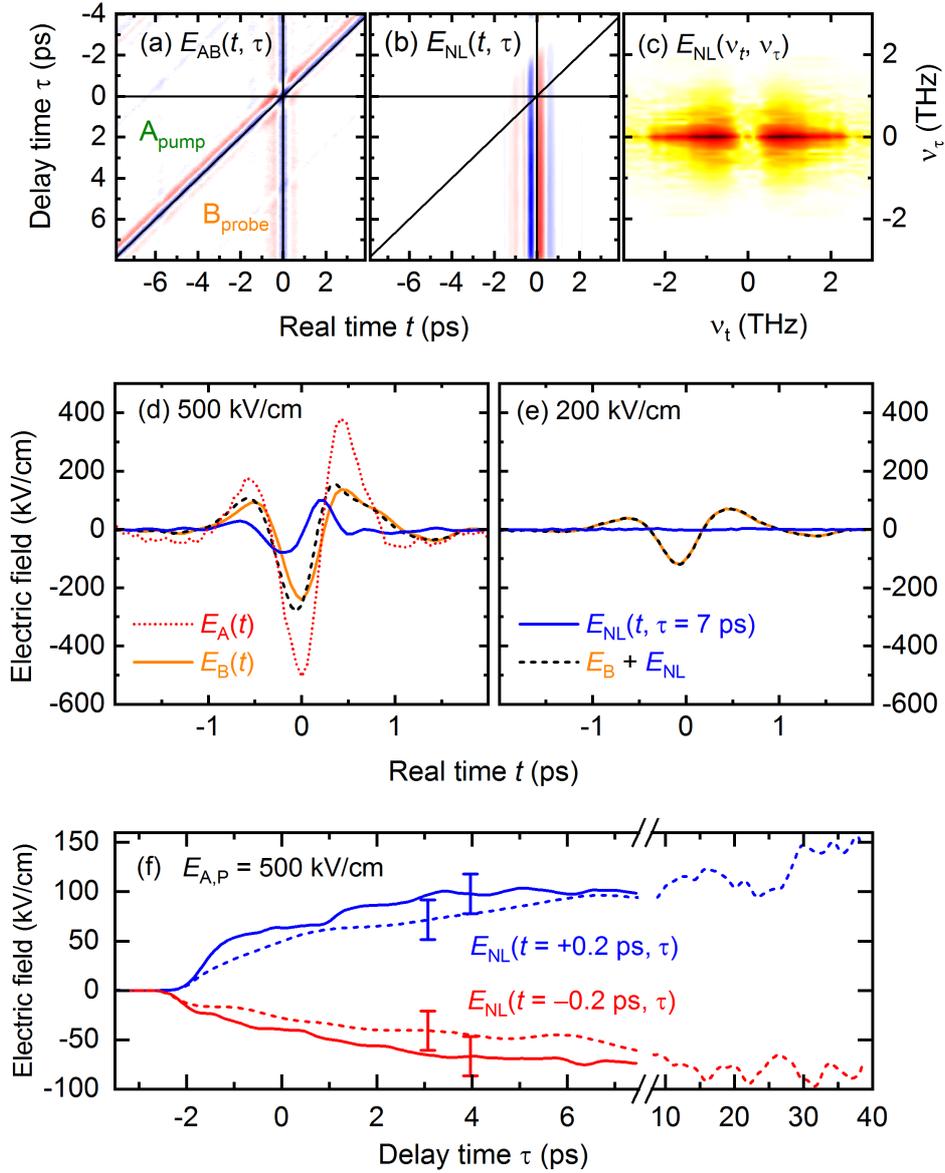

**Figure 2**. Time-resolved nonlinear THz response. (a) Contour plot of the 2D-THz scan $E_{AB}(t,\tau)$ transmitted through the water sample as a function of the real time $t$ and the delay time $\tau$ between pulses A and B. (b) Nonlinear signal field $E_{NL}(t, \tau) = E_{AB}(t, \tau) - E_A(t, \tau) - E_B(t)$ as a function of $t$ and $\tau$. (c) 2D-Fourier transform $E_{NL}(\nu_t,\nu_\tau)$ of $E_{NL}(t, \tau)$ [panel (b)] as a function of detection frequency $\nu_t$ and excitation frequency $\nu_\tau$. (d, e) Nonlinear signal field $E_{NL}(t, \tau = 7\text{ ps})$ (blue line) for a delay time $\tau = 7$ ps together with the probe field $E_B(t)$ (orange line) and the sum $E_{B,T}(t, \tau = 7\text{ ps}) = E_B(t) + E_{NL}(t, \tau = 7\text{ ps})$ (black dashed line) as a function of real time $t$ for peak amplitudes $E_{A,P} = 500$ kV/cm and 200 kV/cm of the incoming pump pulse A. The field $E_{B,T}(t, \tau = 7\text{ps})$ represents the total THz probe field transmitted through the excited sample. The dotted red line in panel (d) shows the transmitted pump field $E_A(t)$. (f) Nonlinear signal field $E_{NL}(t = \pm 0.2\text{ ps}, \tau)$ as a function of delay time $\tau$ for $t = -0.2$ ps (red solid line, signal minimum in Figure 2d) and $t = +0.2$ ps (blue solid line, signal maximum in Figure 2d) for a pump field amplitude $E_{A,P} = 500$ kV/cm. The dashed lines from an independent measurement with a slightly different pump field agree within the experimental accuracy (error bars) and extend to longer delay times $\tau$.

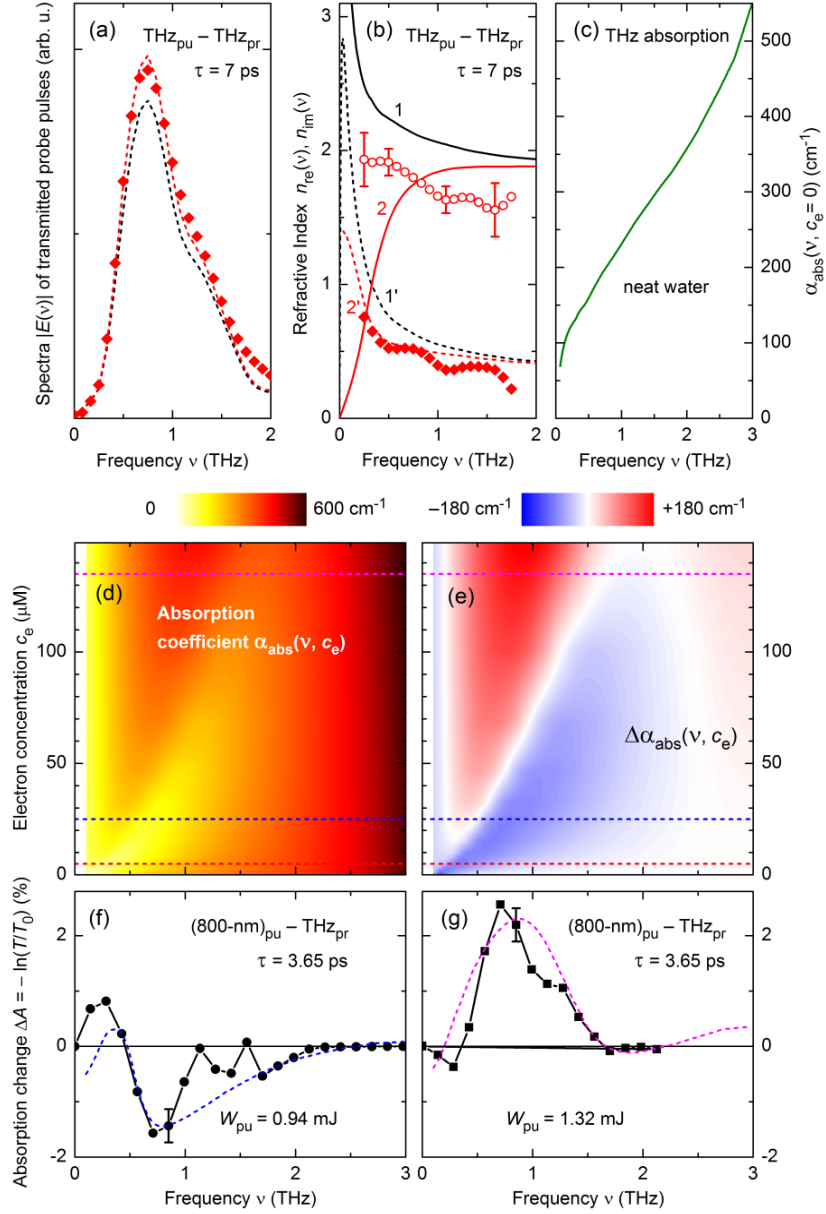

**Figure 3**. Impact of hydrated electrons on the THz dielectric response of water. (a) Transmitted spectrum of the THz probe pulse without excitation (black line) and after pumping by pulse A (symbols, pump-probe delay $\tau = 7$ ps, pump field amplitude $E_{A,P} = 500$ kV/cm). The red line was calculated for water with $c_e = 5$ µM hydrated electrons. (b) Real part $n_{re}(\nu)$ (solid lines) and imaginary part $n_{im}(\nu)$ (dashed lines) of the refractive index $n(\nu)$ in the spectral range from 0 to 2 THz. Black lines (1,1'): neat water. Red lines (2,2'): $n(\nu)$ calculated for $c_e = 5$ µM. The symbols (open circles: $n_{re}(\nu)$, full diamonds: $n_{im}(\nu)$) were derived from the 2D-THz data. (c) THz absorption coefficient $\alpha_{abs}(\nu,c_e) = (4\pi\nu/c_0)n_{im}(\nu,c_e)$ of neat water ($c_e = 0$, $c_0$: velocity of light). (d) Calculated absorption coefficient of water plotted as a function of the THz frequency $\nu$ and electron concentration $c_e$. The dashed lines mark the electron concentrations of panels (a,f,g). (e) Differential absorption coefficient $\Delta\alpha_{abs}(\nu,c_e) = \alpha_{abs}(\nu,c_e) - \alpha_{abs}(\nu,0)$ as a function of $\nu$ and $c_e$. (f,g) Results of pump-probe experiments with 800 nm pump pulses (pulse energy $W_{pu}$) and THz probe pulses. The measured absorption change $\Delta A = -\ln(T/T_0) = \Delta\alpha_{abs}d$ (symbols) is plotted as a function of THz frequency for two pump energies ($T, T_0$: sample transmission with and without pump pulse, $d = 50$ µm: sample thickness). The dashed lines represent differential absorption spectra calculated for (f) $c_e = 24$ µM and (g) $c_e = 85$ µM.

16